\documentclass[RNAAS]{aastex631}
\usepackage{amsmath}
\usepackage{graphicx}

\begin{document}
	
\title{On the plasma quasi-thermal noise in the outer heliosphere}

\correspondingauthor{Nicole Meyer-Vernet}
\email{nicole.meyer@obspm.fr}
\email{alain.lecacheux@obspm.fr}

\author[0000-0001-6449-5274]{Nicole Meyer-Vernet}
\affiliation{LESIA, Observatoire de Paris, PSL Université,\\ CNRS, Sorbonne Université, Université Paris Cité, 92195 Meudon, France}
\author[0009-0005-6822-8997]{Alain Lecacheux}
\affiliation{LESIA, Observatoire de Paris, PSL Université,\\ CNRS, Sorbonne Université, Université Paris Cité, 92195 Meudon, France}

\begin{abstract}
	The recent paper by Li et al. on electron quasi-thermal noise in the outer heliosphere is flawed. It  assumes the plasma drift speed   to be much smaller than the electron thermal speed, even though both quantities are of the same order of magnitude in the outer heliosphere inward of the termination shock, because of the low plasma temperature. In this case, the Langmuir wave dispersion equation  and the quasi-thermal noise in the antenna frame are completely changed.  Furthermore, these calculations neglect the shot noise, which should produce a large contribution  below the plasma frequency with the Voyager antennas in the outer heliosphere.
\end{abstract}


\section{Introduction}
Plasma quasi-thermal noise (QTN) is routinely observed with wave instruments in space, and spectroscopy of this noise is an efficient tool to measure  plasma properties \citep{mey98, mey17}. In weakly magnetised plasmas, the electron QTN consists of a plateau below the plasma frequency $f_p$, mainly produced by electrons crossing the plasma sheath surrounding the antenna, a peak close to $f_p$, mainly produced by electrons of speed close to the large Langmuir wave phase speed, and a power  decreasing as large frequencies \citep{mey89}. This noise has been suggested to be at the origin of a weak  line at the local plasma frequency  discovered in spectra from the Voyager 1 Plasma Wave (PWS) instrument in the very local interstellar medium \citep{ock21,gur21}. \citet{mey22}   have shown that this  line can be explained by the QTN produced by a minute quantity of suprathermal electrons. \citet{mey23} suggested an origin for these  electrons and showed that the original interpretation by \citet{gur21} is questionable. Recently, \citet{li24} published calculations of the QTN in the outer heliosphere on an antenna similar to the Voyager one, using classical expressions of the QTN (\cite{mey17} and references therein) that assume that the plasma drift speed  is much smaller than the electron thermal speed.

\section{The drift speed is not much smaller than the electron thermal speed}

\begin{figure*}
	\centering
	\includegraphics[width=12cm]{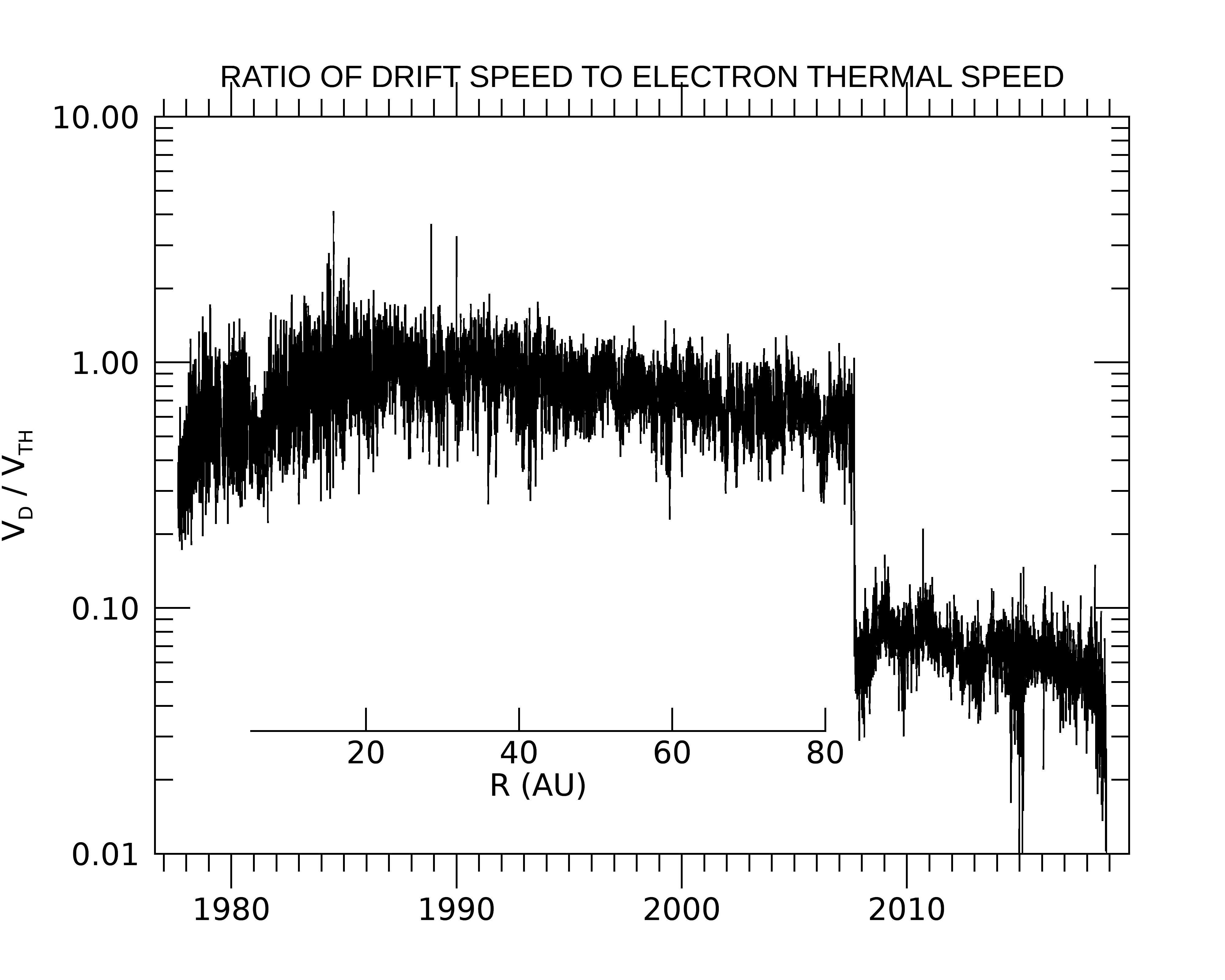}
	\caption{Ratio  of the drift speed  $v_D$ to the electron thermal speed $v_{TH}$ observed by Voyager 2 as a function of time (in years) and heliocentric distance $R$, approximating  the electron temperature by the proton one. The ratio $v_D/v_{TH}$ is close to unity in most of the outer heliosphere, until the crossing of the termination shock where it drops off abruptly.}
	\label{Fig1}
\end{figure*} 

Figure 1 shows the ratio of the drift speed $v_D $   to the electron thermal speed $v_{TH} = (2 k_B T/m)^{1/2}$    measured  by the Plasma (PLS) instrument  \citep{bri77} onboard Voyager 2  in the solar wind. Here  $v_D $ has been estimated from the radial speeds of the solar wind and of the spacecraft, $k_B$ is Boltzmann constant, $m$ is the electron mass, and $T$ is the electron temperature in K, assumed to be close to the proton  one. Note that since the Voyager spacecraft speed of about 15 km/s is much smaller than the solar wind speed, the plasma drift speed relative to the antenna is roughly equal to the solar wind speed. Figure 1 shows that the drift speed is of the same order of magnitude as the electron thermal speed and cannot be neglected in most of the outer heliosphere up to the termination shock, contrary to the assumption made in the calculations  used by \citet{li24}.

The  voltage power spectrum of the plasma quasi-thermal noise at the  
terminals of an  antenna  in a plasma drifting with velocity   ${\bf v_D}$  is 
\begin{eqnarray} 
	V_{\omega }^{2}& = & \frac{2}{\left( 2\pi  \right)^{3}}\int d^{3}k\left| 
	\frac{{\bf k}\cdot {\bf J}}{k} \right|^{2}E^{2}\left( {\bf k},\omega -{\bf 
		k}\cdot {\bf v_{D}} \right)  \protect \label{V}  
\end{eqnarray} 
The first term in the integral in (\ref{V}) involves the antenna response to  electrostatic waves, which depends on the Fourier transform ${\bf J}\left( 
{\bf k} \right)$ of  the current distribution along the antenna.  The second term  is the  Fourier transform of the 
autocorrelation function of the plasma electrostatic field fluctuations in the  antenna frame. Quasi-thermal noise calculations (\cite{mey17} and references therein)  used by \citet{li24}  assume the drift speed $v_D$ to be small enough that $ \omega -{\bf k}\cdot {\bf v_D}$ in (\ref{V}) can be replaced by $ \omega$. Since the   autocorrelation function of the plasma electrostatic field fluctuations is determined by the plasma velocity distributions, the drift speed $v_D$ can be neglected if it is much smaller than the speeds of the particles that determine the QTN, which are mainly the electrons. Since this condition does not hold in most of the outer heliosphere where $v_D$ and $v_{TH}$ are of the same order of magnitude, the QTN will be changed by a large amount. In extreme cases when $v_D>v_{TH} $, the resistance of the antenna could even become negative, producing unreliable electric field measurements \citep{mey89bis}.

\section{The shot noise is not negligible in the outer heliosphere \\ with the Voyager antenna}

A further problem arises in the calculations published by \citet{li24}. In the solar wind, the   photoelectron flux emitted by electric antennas  largely exceeds the collected plasma electron flux, making the floating potential positive \citep{mey07}. This holds in most of the outer heliosphere, where the electron flux decreases with heliocentric distance slightly faster than the photoelectric emission. The shot noise on the Voyager antennas, of radius $a$ and length $L$ much smaller than the plasma Debye length,  would thus be approximately,  if  $v_D \ll v_e$  as wrongly assumed by \citet{li24}
\begin{equation}
	V_{\mathrm{shot}}^2 \simeq 2 \times 10^{-16} T^{1/2} (a/L) [\ln(L/a)-1]^2 (f_p/f)^2  \label{VI}
\end{equation}
in S.I. units, for $f<f_p$  \citep{mey89}. Equation (\ref{VI}), which holds in a maxwellian plasma, is weakly affected by the presence of suprathermal electrons \citep{mey17}. However,  the small temperature  in the outer heliosphere implies  also that the antenna positive potential, equal to a few times the photoelectron temperature in electron volts (eV), can become larger than the electron temperature. Both the antenna potential and the drift speed will increase the shot noise, which is   proportional to the electron flux impacting the antennas. With  the Voyager antennas of radius $a = 0.635$ cm \citep{sca77}, this yields $	V_{\mathrm{shot}}^2 > 10^{-15} (f_p/f)^2 \; T_{eV}^{1/2} $ V$^2/$Hz, where $T_{eV}$ is the temperature in eV. This noise should thus exceed the QTN plateau plotted by  \citet{li24} in most of the outer heliosphere, except with antennas of radius much smaller than on Voyager  or being negatively biased \citep{mey17}.

\section{Conclusion}

The QTN spectra published by \citet{li24} are invalid in most of the outer heliosphere up to the termination shock, since the calculations assume that the drift speed is negligible compared to the electron thermal speed. Furthermore, the shot noise was neglected in this paper, even though it is significant on Voyager antennas in the heliosphere, and would require much thinner antennas and/or a negative biasing to become negligible.

\vspace{1cm}

The PLS data could be obtained from the  site: \url{ftp://space.mit.edu/pub/
plasma/vgr/v2}.

\end{document}